\documentclass[a4paper,fleqn, usenatbib]{mnras}

\usepackage[T1]{fontenc}
\usepackage{ae,aecompl}

\usepackage{graphicx}	% Including figure files
\usepackage{amsmath}	% Advanced maths commands
\usepackage{amssymb}	% Extra maths symbols

\title[Dust reverberation mapping of H0507+164]{Determination of the size of the dust torus 
in H0507+164 through optical and infrared monitoring}

\author[Mandal et al.]{Amit Kumar Mandal$^{1,2}$\thanks{E-mail: amitkumar@iiap.res.in},
Suvendu Rakshit$^{1, 3}$\thanks{E-mail: suvenduat@gmail.com},
Kshama S. Kurian$^{1}$,
C. S. Stalin$^{1}$\thanks{E-mail: stalin@iiap.res.in},
\newauthor
Blesson Mathew$^{2}$, 
Sebastian Hoenig$^{4}$,
Poshak Gandhi$^{4}$,
Ram Sagar$^{1}$
and \newauthor
M. B. Pandge$^{5}$\thanks {DST INSPIRE Faculty}
\\\\
$^{1}$Indian Institute of Astrophysics, Block II, Koramangala, Bangalore 560 034, India\\
$^{2}$Department of Physics, Christ University, Hosur Road, Bangalore 560 029, India\\
$^{3}$Astronomy Program, Department of Physics and Astronomy, Seoul National University, Seoul 151-742, Republic of Korea \\
$^{4}$Department of Physics \& Astronomy, University of Southampton, Southampton SO17 1BJ, UK\\
$^{5}$Dayanand Science College, Barshi Road, Latur, Maharashtra 413512, India
}

\date{Accepted 2018 January 19. Received 2018 January 11; in original form 2017 September 5}

\pubyear{2018}

\begin{document}
\label{firstpage}
\pagerange{\pageref{firstpage}--\pageref{lastpage}}
\maketitle

\begin{abstract}

The time delay between flux variations in different wavelength
bands can be used to probe the inner regions of active galactic
nuclei (AGN). Here, we present the first measurements of the 
time delay between optical
and near-infrared (NIR)  flux variations
in H0507+164, a nearby Seyfert 1.5 galaxy
at $z$ = 0.018. The observations in the 
optical $V$-band and NIR $J$, $H$ and $K_s$ bands
carried over 35 epochs during the period October 2016 to April 2017
were used to
estimate the inner radius of the dusty torus.  From a
careful reduction and analysis of the data using cross-correlation
techniques,  we found delayed responses of
the $J$, $H$ and $K_s$ light curves to the $V$-band light curve. In the rest
frame of the source, the lags between optical and NIR bands are found to
be $27.1^{+13.5}_{-12.0} \, \mathrm{days}$ ($V$ vs. $J$),
$30.4^{+13.9}_{-12.0} \, \mathrm{days}$ ($V$ vs. $H$) and
$34.6^{+12.1}_{-9.6} \, \mathrm{days}$ ($V$ vs. $K_s$). The lags
between the optical and different NIR bands are thus consistent with each 
other. 
The measured lags
indicate that the inner edge of dust torus is located at a distance
of  0.029 pc from the central UV/optical AGN continuum. This is larger
than the radius of the broad line region of this object determined
from spectroscopic monitoring observations thereby supporting the
unification model of AGN.
The location of H0507+164 in the $\tau$ - $M_V$ plane indicates that our results are in excellent agreement with the now known
lag-luminosity scaling relationship for dust in AGN.
\end{abstract}

\begin{keywords}
galaxies: active $-$ galaxies: Seyfert $-$ (galaxies:) quasars: individual (H0507+164)
\end{keywords}

%%%%%%%%%%%%%%%%%%%%%%%%%%%%%%%%%%%%%%%%%%%%%%%%%%

%%%%%%%%%%%%%%%%% BODY OF PAPER %%%%%%%%%%%%%%%%%%

\section{Introduction}

Active Galactic Nuclei (AGN) are among the most luminous objects in the
Universe that produce very high luminosity in a very concentrated volume,
powered by the accretion of
matter onto super-massive black hole (SMBH) located at the centers of galaxies.
According to the unified model of AGN, a dusty torus that surrounds
the central SMBH, the accretion disk and the broad line region (BLR)
play a key role in the identification of Seyfert galaxies, a category of 
AGN, into Seyfert 1 and
Seyfert 2 galaxies \citep{1993ARA&A..31..473A}. Evidence for their presence is seen in the
broad band spectral energy distribution (SED) of AGN, such as the
presence of the big blue bump
in the optical-UV wavelength region, which is a signature of the
accretion disk \citep{1987ApJ...321..305C,
1982ApJ...254...22M,1978Natur.272..706S} and a bump in the
IR region \citep{1989ApJ...347...29S,1987ApJ...320..537B,
1993ApJ...404...94K} which is a signature of the presence of the dusty torus.

The torus, which obscures the central engine, is the dominant source of IR
radiation in most of the AGN.
It is not easy to spatially resolve the components of the AGN such as the BLR and torus by any current
imaging techniques. The IR radiation of AGN coming from the torus that extends
in size from sub-pc to pc size scales 
\citep{2011A&A...527A.121K} could in principle be resolvable
using IR interferometric observations. However, such interferometric 
observations as of today are in a minority \citep{2011A&A...536A..78K,2013A&A...558A.149B} and limited by the current diameters of 
optical/IR telescopes. 
Alternatively, the extent and nature of the torus can be studied
via the technique of reverberation mapping \citep{1982ApJ...255..419B}.
This technique relies on the flux variability of AGN, which has been known
since their discovery.  The continuum emission from AGN is believed to arise
from the accretion disk and it is expected to vary over a range of time scales
from hours to days. The
variations in the continuum flux are observed at a later time in the fluxes of their broad emission lines. Similarly, there is also
delayed response of the IR continuum emission (from the torus) to the changes
in the optical continuum from the central AGN, which indicates that
the NIR continuum and UV/optical continuum emission are
causally connected.  This delay between
the optical and IR variation when measured can give the radius of the inner most hot dust
torus $r_{\mathrm{dust}}$ = $\tau \times c$, where c is the speed of light and
$\tau$ is the time delay between optical and IR variations. This method of measuring the delayed response of the IR emission relative to
the optical emission and consequently determine the size of the torus
is referred to as dust reverberation mapping (DRM). 
%In addition to providing
%the size of the torus, DRM also allows one to study dust morphology at 
%different wavelengths. %and can help to understand accretion and wind components linked to the accretion state.

Studies of DRM in few Seyfert galaxies 
\citep{2004ApJ...600L..35M,2006ApJ...639...46S,2014ApJ...788..159K} have led to 
the establishment of a correlation between the inner radius of the dust torus 
and the UV luminosity as $r_{\mathrm{dust}}$ $\varpropto$ $L^{0.5}$ 
\citep{2007A&A...476..713K}. Also,  
\cite{2007A&A...476..713K} established a relation between the sublimation 
radius ($R_{\mathrm{sub}}$) 
which is defined as the radius at which dust particle sublimates and the dust grain size ($a$) 
as 
\begin{align}
R_{\mathrm{sub}} = 1.3 \left(\frac{L_{UV}}{10^{46} \mathrm{erg s^{-1}}}\right)^{0.5} \left(\frac{T_{\mathrm{sub}}}{1500 K}\right)^{-2.8} \left(\frac{a}{0.05\mu \mathrm{m}}\right)^{-0.5} \mathrm{pc}
\label{eq:1}
\end{align}
Considering sublimation temperature ($T_{\mathrm{sub}}$) to be the dust 
temperature 
$T_{\mathrm{dust}}$= 1700 K which was evaluated from the NIR colors of the 
variable flux component for Seyfert 1 galaxies  \citep{2006ApJ...652L..13T,
2014ApJ...788..159K} and grain size $a = 0.1 \mu$m, the sublimation radius 
from equation \ref{eq:1} can be written as log $R_{\mathrm{sub}}$/pc = 
$-0.80 + 0.5 \log(L_V/10^{44}$ erg s$^{-1}$), 
where $L_{UV} = 6 \,L_V$ \citep{2007A&A...476..713K}. This is in 
excellent agreement with the relation of log $r_{\mathrm{dust}}$/pc =
$-0.88 + 0.5 \log(L_V/10^{44}$ erg s$^{-1}$) obtained by a linear regression
fit to the reverberation mapping data of a few Seyfert galaxies by  
\cite{2014ApJ...788..159K}. Thus, DRM observations using the $K$-band in the infrared region can define 
the inner radius of the dust torus for an AGN.

The emission line lag  \citep[e.g., ][]{2011ApJ...740L..49W,2011A&A...535A..73H,
2013A&A...556A..97C} and dust reverberation lag \citep[e.g.,][]{1999AstL...25..483O,
2001ASPC..224..149O,2014ApJ...784L...4H,2014ApJ...784L..11Y,
2017MNRAS.464.1693H} have been proposed to serve as standard candles. However, 
the later shows a stronger correlation between lag and luminosity than former 
and needs only photometric monitoring making it preferable for standard candle 
although the emission line lags can be measured up to $z\sim 4$ 
\citep[see][]{2011ApJ...740L..49W}. Unfortunately, only handful number of 
objects has dust lag measurement \citep{2006ApJ...639...46S,
2014A&A...561L...8P,2014ApJ...788..159K}, which motivated 
\citet{2017MNRAS.464.1693H} to carry out a large dust reverberation mapping 
(DRM) program, ``VEILS'' (VISTA Extragalactic Infrared Legacy Survey) to use 
dust lag as standard candle for cosmology and constrain cosmological parameters.
The ``VEILS'' survey will target about 1350 Seyfert 1 galaxies 
in the redshift range 
$0.1<z<1.2$, however, missing the objects in the local universe which are 
crucial in determining the normalization parameter of the AGN distance moduli 
\citep[see][]{2017MNRAS.464.1693H}. Thus, dust reverberation mapping of local 
AGN will not only allow to estimate the extent of the torus, dust morphology 
and verify orientation 
dependent unification model but also help to estimate the normalization constant of the Hubble function.

To complement the VEILS program, in the nearby Universe, we are carrying out
monitoring observations of a carefully selected sample of AGN with
redshift less than 0.1 as part of our long term project called
REMAP (REverberation Mapping of AGN Program). Observations towards
this project has been carried out using the 2 m Himalayan Chandra Telescope at Hanle, India.
Such a systematic study in conjunction with VEILS will serve as an excellent 
database to probe the complex interplay between optical continuum
and dust emission in AGN. Here, we present the results of DRM 
monitoring of one local ($z=0.018$) Seyfert 
1.5 galaxy H0507+164. 
The structure of this paper is as follows. 
In section \ref{sec:observation} we briefly describe the observation and data 
reduction processes. The analysis is presented in section 
\ref{sec:Analysis}. The results are discussed in   
section \ref{sec:discussion} followed by the summary in section 
\ref{sec:summary}. For the cosmological parameters we assumed $H_0=73$ 
km $\mathrm{s^{-1}Mpc^{-1}}$, $\Omega_m=0.27$ and $\Omega_{\lambda}=0.73$ 
\citep{2014ApJ...788..159K}.        

\section{Observation and data reduction}\label{sec:observation}

\subsection{The Sample}
The sample of sources for our DRM project were drawn from 
\cite{2015PASP..127...67B} that have BLR lag from spectroscopic reverberation 
observations. From this list, we have selected only those sources
that are accessible for observations using the 2m Himalayan Chandra
Telescope (HCT) located at Hanle and operated by the Indian Institute of 
Astrophysics\footnote{\url{http://www.iiap.res.in/centres/iao}}.
In this paper, we present the result of the first source monitored 
for  DRM using the HCT, namely H0507+164, a local Seyfert 1.5 galaxy at 
$z$ = 0.018 with RA = 05:10:45.5 and DEC = +16:29:56. It is bright
with an optical $g$ band magnitude of 10 mag from the SIMBAD database\footnote{\url{http://simbad.u-strasbg.fr/simbad/}}. It has a black hole mass of $9.62^{+0.33}_{-3.73}\times 10^6 M_{\odot}$ deduced from spectroscopic reverberation observations \citep{2011MNRAS.416..225S}. It is detected in the NRAO/VLA Sky Survey (NVSS\footnote{\url{http://www.cv.nrao.edu/nvss/NVSSlist.shtml}}) with a flux density of $6\pm0.5$ mJy at 1.4 GHz. However, it is radio quiet with a radio loudness parameter, $R$ (ratio of flux density in the radio-band at 1.4 GHz to the optical $V$-band flux density) of $2.26\pm0.26$.

\subsection{Observations}
The observations were carried out for a total of 35 epochs during the 
period October 2016 to April 2017  using HCT 
The telescope is a Ritchey-Chretien system with an f/9 beam. The optical observations were
carried out using the Himalayan Faint Object Spectrograph and Camera (HFOSC) mounted at the Cassegrain focus and equipped with 
a 2K $\times$ 4K SiTe CCD system. This CCD  has a readout 
noise and gain of 4.8 electrons/ADU and 1.22 electrons respectively. The observations
were carried out in binned mode using only the central 2K $\times$ 2K 
region. Each pixel of 
the CCD in this mode corresponds to 0.3 $\times$ 0.3 arcsec$^2$, 
and for the 
imaging observations reported here has a field of view of 10  
$\times$ 10 arcmin$^2$. The typical
exposure time in $V$-band is about 50 seconds. The IR observations   
in $J$, $H$ and $K_s$ bands were done subsequently to the $V$-band observations at each 
epoch 
were carried out using the TIFR Near Infrared Spectrometer (TIRSPEC)
mounted on one of the side ports of HCT
\citep{2014JAI.....350006N}. The detector used in TIRSPEC 
is a 1024 $\times$ 1024 HgCdTe array with a pixel size of 18 $\mu$m covering a field of view of 5 $\times$ 5 arcmin$^2$.  
It has a readout noise and gain of 25 electrons and 6 electrons/ADU 
respectively. The IR 
observations were performed in 
dithered mode consisted of five exposures each of 20 sec 
at three dither positions  for each of three IR filters namely $J$, $H$ and $K_s$.
Apart from the science frames, sky regions were also observed in the same
dithering pattern as the object to generate master sky frame.

\subsection{Data reduction}

\begin{figure}
\begin{center}
\includegraphics[scale=0.25]{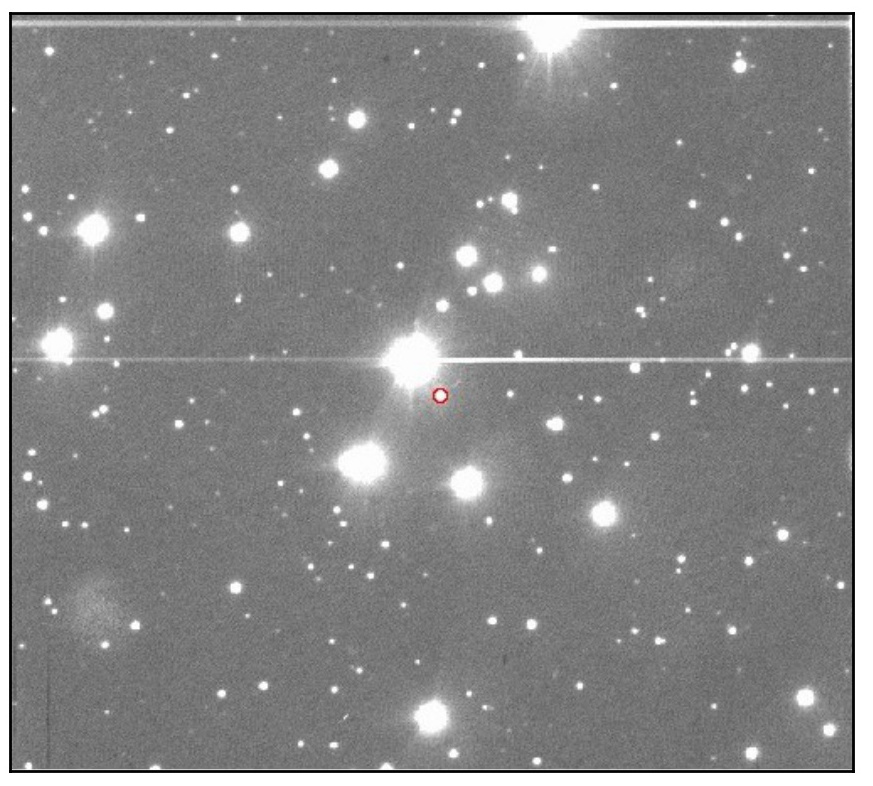}
\end{center}
\caption{Observed $V$-band image of H0507+164. The integration time
is 50 seconds and each side is 10 arcmin. The target source is shown 
with a circle.}
\label{Fig:FOV}
\end{figure}

The optical data reduction was done using IRAF (Image Reduction and 
Analysis Facility\footnote{IRAF is operated by the Association of Universities 
for Research in Astronomy, Inc., under cooperative agreement with the National 
Science Foundation.}) and MIDAS (Munich Data Analysis System\footnote{MIDAS is 
the trade-mark of the European Southern Observatory}). For image reduction, we 
followed the standard procedure, such as bias $\&$ dark subtraction and flat-fielding. 
After removing the cosmic rays, the optical frames were aligned and then 
combined using {\tt imalign} and {\tt imcombine} tasks in IRAF for each day. 
Point spread function (PSF) 
photometry was carried out on the combined images using the 
{\tt daophot} and {\tt allstar} 
packages in MIDAS to find the instrumental magnitudes of the object and the 
comparison stars present in the CCD frames. The observed $V$-band image is shown in Figure \ref{Fig:FOV}.

Reduction of the NIR images was performed using TIRSPEC NIR Data 
Reduction Pipeline \citep{2014JAI.....350006N} for each band. The 
pipeline subtracts the dark obtained for the same exposure time as the 
science exposures and then uses twilight flats for flat fielding and produce final combined images. PSF photometry on the combined images 
was carried out by using MIDAS to get the instrumental magnitudes.

The object appears as a point source in our optical and IR data, thereby, making it difficult to model the host
galaxy from the observed data and remove its contribution to the measured brightness. Therefore, subtraction of the host galaxy contribution to the
measured brightness of the source was not carried out. Also, we have not corrected the observed flux values for galactic 
extinction. These effects will be small and a constant fraction of the measured flux values, thus having negligible 
effects on the cross-correlation analysis.

\begin{figure}
\begin{center}
\resizebox{8cm}{12cm}{\includegraphics{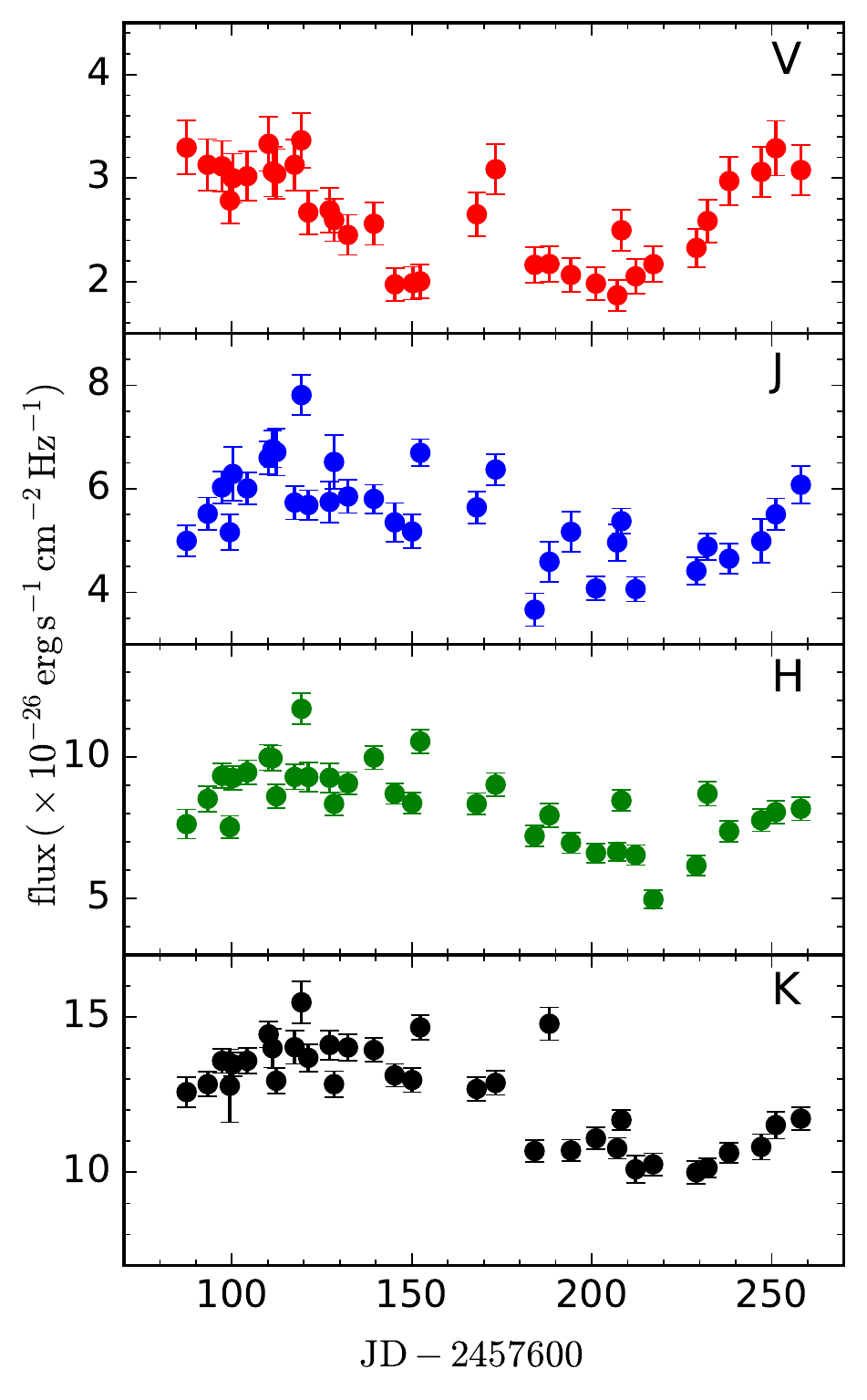}}
\end{center}
\caption{The light curves of H0507+164 in  $V$, $J$, $H$, $Ks$-bands 
for the period from October 2016 to February 2017. The NIR light curves were
corrected for the contamination of emission from the accretion disk.}
\label{fig:lc}
\end{figure}

\begin{center}
\begin{table*}
\caption{Log of observations and results of photometry. Here JD is in days, the 
fluxes in different bands and their associated errors are in units
of 10$^{-26}$ erg s$^{-1}$ cm$^{-2}$ Hz$^{-1}$.}
\begin{tabular}{|cccccccccccc|} \hline
JD$_V$          &  $F_V$     & $\sigma_V$ & JD$_J$      & $F_J$    & $\sigma_J$ &  JD$_H$   & $F_H$   & $\sigma_H$ & JD$_{K_s}$ &   $F_{K_s}$   & $\sigma_{K_s}$         \\ \hline
2457687.4272  & 3.295  & 0.260 & 2457687.4429 &  4.994 & 0.302 & 2457687.4374 &  7.628 &  0.517 & 2457687.4316  & 12.582 &  0.490 \\
2457693.2663  & 3.129  & 0.247 & 2457693.2765 &  5.519 & 0.312 & 2457693.2819 &  8.521 &  0.457 & 2457693.2875  & 12.835 &  0.399 \\
2457697.2763  & 3.115  & 0.246 & 2457697.2802 &  6.027 & 0.313 & 2457697.2858 &  9.339 &  0.432 & 2457697.2950  & 13.585 &  0.387 \\
2457699.4455  & 2.784  & 0.221 & 2457699.4277 &  5.161 & 0.346 & 2457699.4219 &  7.520 &  0.399 & 2457699.4164  & 12.783 &  1.174 \\
2457700.2714  & 2.999  & 0.238 & 2457700.2536 &  6.287 & 0.521 & 2457700.2473 &  9.262 &  0.429 & 2457700.2416  & 13.473 &  0.386 \\
2457704.2750  & 3.019  & 0.238 & 2457704.2595 &  6.008 & 0.305 & 2457704.2541 &  9.459 &  0.428 & 2457704.2486  & 13.589 &  0.403 \\
2457710.2058  & 3.331  & 0.262 & 2457710.2274 &  6.596 & 0.317 & 2457710.2219 &  9.989 &  0.455 & 2457710.2165  & 14.439 &  0.422 \\
2457711.3872  & 3.064  & 0.241 & 2457711.3110 &  6.768 & 0.357 & 2457711.3051 &  9.962 &  0.447 & 2457711.2986  & 13.995 &  0.621 \\
2457712.3179  & 3.041  & 0.240 & 2457712.3339 &  6.710 & 0.453 & 2457712.3279 &  8.612 &  0.415 & 2457712.3219  & 12.946 &  0.416 \\
2457717.4375  & 3.129  & 0.247 & 2457717.4540 &  5.734 & 0.324 & 2457717.4488 &  9.297 &  0.441 & 2457717.4435  & 14.027 &  0.536 \\
2457719.2792  & 3.365  & 0.265 & 2457719.3544 &  7.811 & 0.384 & 2457719.3489 & 10.171 &  0.543 & 2457719.3433  & 15.470 &  0.668 \\
2457721.2253  & 2.668  & 0.212 & 2457721.2420 &  5.685 & 0.289 & 2457721.2364 &  9.295 &  0.526 & 2457721.2308  & 13.682 &  0.440 \\
2457727.1733  & 2.690  & 0.213 & 2457727.1922 &  5.747 & 0.394 & 2457727.1868 &  9.270 &  0.518 & 2457727.1812  & 14.099 &  0.472 \\
2457728.4147  & 2.596  & 0.206 & 2457728.3990 &  6.519 & 0.523 & 2457728.3928 &  8.346 &  0.411 & 2457728.3874  & 12.835 &  0.422 \\
2457732.2453  & 2.451  & 0.194 & 2457732.2619 &  5.852 & 0.318 & 2457732.2559 &  9.073 &  0.401 & 2457732.2499  & 14.018 &  0.423 \\
2457739.4652  & 2.560  & 0.205 & 2457739.4483 &  5.804 & 0.276 & 2457739.4421 &  9.983 &  0.415 & 2457739.4363  & 13.936 &  0.382 \\
2457745.2972  & 1.974  & 0.158 & 2457745.2543 &  5.352 & 0.374 & 2457745.2485 &  8.704 &  0.364 & 2457745.2420  & 13.121 &  0.365 \\
2457750.3023  & 1.985  & 0.158 & 2457750.0752 &  5.175 & 0.327 & 2457750.0811 &  8.368 &  0.371 & 2457750.0890  & 12.969 &  0.394 \\
2457752.3102  & 2.004  & 0.162 & 2457752.3535 &  6.695 & 0.262 & 2457752.3476 & 10.556 &  0.411 & 2457752.3415  & 14.664 &  0.403 \\
2457768.0595  & 2.651  & 0.213 & 2457768.0408 &  5.642 & 0.307 & 2457768.0331 &  8.345 &  0.385 & 2457768.0276  & 12.681 &  0.376 \\
2457773.3053  & 3.086  & 0.243 & 2457773.2892 &  6.372 & 0.302 & 2457773.2830 &  9.025 &  0.413 & 2457773.2769  & 12.876 &  0.389 \\
2457784.1059  & 2.163  & 0.173 & 2457784.1302 &  3.666 & 0.316 & 2457784.1220 &  7.208 &  0.372 & 2457784.1136  & 10.684 &  0.355 \\
2457788.2195  & 2.171  & 0.173 & 2457788.2422 &  4.591 & 0.392 & 2457788.2362 &  7.938 &  0.422 & 2457788.2302  & 14.780 &  0.528 \\
2457794.2058  & 2.065  & 0.164 & 2457794.2221 &  5.169 & 0.386 & 2457794.2160 &  6.960 &  0.363 & 2457794.2104  & 10.702 &  0.350 \\
2457801.1541  & 1.982  & 0.158 & 2457801.1683 &  4.076 & 0.226 & 2457801.1626 &  6.604 &  0.338 & 2457801.1568  & 11.091 &  0.356 \\
2457807.0827  & 1.868  & 0.149 & 2457807.0662 &  4.963 & 0.350 & 2457807.0598 &  6.646 &  0.314 & 2457807.0538  & 10.771 &  0.339 \\
2457808.2595  & 2.497  & 0.197 & 2457808.2429 &  5.374 & 0.244 & 2457808.2372 &  8.461 &  0.369 & 2457808.2315  & 11.685 &  0.324 \\
2457812.2031  & 2.052  & 0.166 & 2457812.1793 &  4.063 & 0.238 & 2457812.1733 &  6.533 &  0.356 & 2457812.1673  & 10.096 &  0.439 \\
2457817.1344  & 2.169  & 0.173 &  ---         & ---    & ---   & 2457817.1214 &  4.961 &  0.317 & 2457817.0465  & 10.255 &  0.356 \\
2457829.0926  & 2.326  & 0.186 & 2457829.0757 &  4.414 & 0.261 & 2457829.0698 &  6.161 &  0.344 & 2457829.0639  &  9.997 &  0.359 \\
2457832.1700  & 2.586  & 0.206 & 2457832.1545 &  4.882 & 0.251 & 2457832.1483 &  8.705 &  0.425 & 2457832.1470  & 10.143 &  0.306 \\
2457838.1930  & 2.972  & 0.235 & 2457838.1716 &  4.650 & 0.287 & 2457838.1656 &  7.371 &  0.374 & 2457838.1601  & 10.626 &  0.333 \\
2457847.1147  & 3.061  & 0.243 & 2457847.1010 &  4.989 & 0.422 & 2457847.0918 &  7.763 &  0.389 & 2457847.0861  & 10.814 &  0.411 \\
2457851.1541  & 3.289  & 0.265 & 2457851.1399 &  5.507 & 0.306 & 2457851.1344 &  8.052 &  0.399 & 2457851.1289  & 11.519 &  0.436 \\
2457858.1201  & 3.078  & 0.244 & 2457858.1044 &  6.078 & 0.356 & 2457858.0989 &  8.170 &  0.414 & 2457858.0933  & 11.727 &  0.373 \\
\hline
\end{tabular}
\label{table2}
\end{table*}
\end{center}

\subsection{Subtraction of the accretion disk component from the NIR bands}

The observed NIR fluxes are dominated by emission from  the torus due to  
re-processing 
of the central UV/optical emission by the hot dust. However, it
is contaminated by variable flux from the accretion disk component \citep{2006ApJ...652L..13T,
2008Natur.454..492K,2011MNRAS.415.1290L}. This would make the time lag 
calculated by cross correlation analysis  between the optical and IR light 
curves shorter than the actual lag of the dust-torus emission \citep{2014ApJ...788..159K}. Thus, to get the true time lag, 
between optical and NIR flux variations, the contribution of accretion disk to 
the measured NIR fluxes needs to be removed.  We estimated the contribution of 
the accretion disk in the $J$, $H$, $Ks$-bands in  each of 
the epochs of observations  by considering a power-law 
spectrum of the accretion disk following   
\citet{2014ApJ...788..159K} and given by
\begin{equation}
f_{\mathrm{NIR,disk}}(t) = f_{V}(t)\left(\frac{\nu_{\mathrm{NIR}} }{\nu_V}\right)^{\alpha_{\nu}}
\label{eq:nir_eq}
\end{equation}
where, $f_{V}(t)$ is the $V$-band flux at time `$t$', $\nu_V$ and 
$\nu_{\mathrm{NIR}}$ are effective frequencies of $V$-band and 
NIR bands respectively and  $\alpha_{\nu}$ is the power-law index.
At any given epoch, the observations in the optical and each of the 
NIR bands differ by less than 300 seconds, and therefore,
for the purpose of removing the contribution of accretion disk to each of the 
NIR bands, they were considered as simultaneous observations.
To calculate $f_{\mathrm{NIR,disk}}$($t$), we assumed power-law index 
$\alpha_{\nu}$ to be equal to 1/3 following \citet{2014ApJ...788..159K}. AGN do show spectral variability with a bluer-when-brighter behavior \citep{2011A&A...525A..37M} which demands adoption of 
a time dependent $\alpha_{\nu}$  to correct for the contribution of
accretion disk emission to NIR. Our single optical $V$-band observations hinder
determination of $\alpha_{\nu}$ for each epoch of our observations. Use of
single $\alpha_{\nu}$ might have some effect on the corrected $J$-band fluxes,
however, it will have negligible effect on the corrected $K_s$-band flux values. 
 
Prior to the calculation and subsequent subtraction of the accretion disk 
component from the observed NIR fluxes, the observed
instrumental magnitudes were converted to apparent magnitudes via 
differential photometry of few stars in the field whose apparent magnitudes 
were taken from the  SIMBAD database. The apparent magnitudes were then 
corrected for Galactic extinction taken from the NASA/IPAC Extragalactic 
database (NED\footnote{\url{https://ned.ipac.caltech.edu/}}). These magnitudes were then converted into 
fluxes and the contribution of accretion disk 
evaluated using equation \ref{eq:nir_eq} were subtracted. The final fluxes thus obtained were 
used to generate optical and NIR light curves for 
further analysis. The errors in the final flux values in different filters  were obtained through 
propagation of errors during the various steps of flux determination.

\section{Analysis}\label{sec:Analysis}

\subsection{Light curves}
The light curves of H0507+164 in optical $V$-band and NIR $J$, $H$ and $K_s$ 
bands are shown in Figure \ref{fig:lc}. As the source is radio-quiet, not much contamination to the observed flux variations is expected from the jet of the source. The variability nature of the source was 
characterized using the normalized excess variance 
\citep{2002ApJ...568..610E,2003MNRAS.345.1271V,2017MNRAS.466.3309R} defined 
as 
\begin{equation}
 F_{\mathrm{var}}  =  \frac{\sqrt{S^2-\delta^2}}{\langle f \rangle} \\
\end{equation}

Here $\langle f \rangle$,  $S^2$ , $\delta$ are the mean flux, variance and mean error respectively for N observations and are given as
\begin{gather}
\langle f \rangle   =  \frac{1}{N}\sum_{i=1}^{N}f_i \\
S^2  =  \frac{1}{N-1} \sum_{i=1}^{N}(f_i - \langle f \rangle)^2 \\
 \delta^2  = \frac{1}{N} \sum_{i=1}^{N}\delta_i^2, \\
\end{gather}
where  $f_i$ and $\delta_i$ are the flux and error for the $i^{th}$ measurement respectively. The uncertainties in the $F_{\rm{var}}$ are evaluated following \citet{2002ApJ...568..610E} and given as
\begin{gather}
\mathrm{err}(F_{\mathrm{var}})  =  \sqrt{ \bigg(\sqrt{\frac{1}{2N}}\frac{\delta^2}{\langle f \rangle^2F_{\mathrm{var}}}\bigg)^2 + \bigg(\sqrt{\frac{\delta^2}{N}}\frac{1}{\langle f \rangle}\bigg)^2}
\end{gather}
In addition to calculating $F_{\mathrm{var}}$ we also calculated the
ratio $R_{\mathrm{max}}$ between the maximum and minimum flux in the light curves. 
The results of variability analysis are given in Table \ref{table:var}.
It is noticed that $F_{\mathrm{var}}$ in the optical $V$-band and NIR J 
and H-bands are similar, while the $F_{\mathrm{var}}$ in $K_s$ band is lower 
than that in the other bands. An  anti-corelation  between 
variability amplitude and wavelength has been found recently based on 
analysis of NIR variability of a large sample of AGN by 
\cite{2017ApJ...849..110S}. Also, compared to the 
optical bands, amplitude of NIR variation is expected to be 
smaller \citep{2002ApJS..141...45E}. Though our data do not reveal a clear
anti-correlation of variability amplitude with wavelength as found recently
\citep{2017ApJ...849..110S}, the $F_{\mathrm{var}}$ in the NIR bands are
lower than of the optical V-band as expected \citep{2002ApJS..141...45E}.
Therefore, the low amplitude of variations seen in the $K_s$ band do 
not point to any inherent bias in 
the $K_s$ band data.

\begin{center}
 \begin{table}
\caption{Variability statistics in $VJHK_s$ bands in observer's frame. Here, $\lambda_{\mathrm{eff}}$ is the 
effective wavelength in Angstroms. The average values ($\langle f \rangle$) and the standard deviation $\sigma$ of the $VJHK_s$ light curves} are in units of 10$^{-26}$ erg s$^{-1}$ cm$^{-2}$ Hz$^{-1}$.
%\hskip-2.0cm
\begin{tabular}{|c|c|c|c|c|c|}
\hline
Filter & $\lambda_{\mathrm{eff}}$& $ \langle f \rangle$ & $\sigma$   & $F_{\mathrm{var}}$ & $R_{\mathrm{max}}$   \\
   &   &              &           &   &\\
 \hline
$V$  & 5510 & 2.66 & 0.48  & 0.159$\pm0.005$ & 1.801 \\
$J$  & 12200 & 5.56 & 0.89 & 0.149$\pm0.003$ & 2.131 \\
$H$  & 16300 & 8.39 & 1.36 & 0.154$\pm0.002$  & 2.360 \\
$K_s$ & 21900 & 12.56 & 1.56 & 0.119$\pm0.001$ & 1.548 \\ \hline
\end{tabular}
\label{table:var}
\end{table}
\end{center}

\subsection{Cross-Correlation Analysis}
\begin{figure}
\begin{center}
\resizebox{8cm}{14cm}{\includegraphics{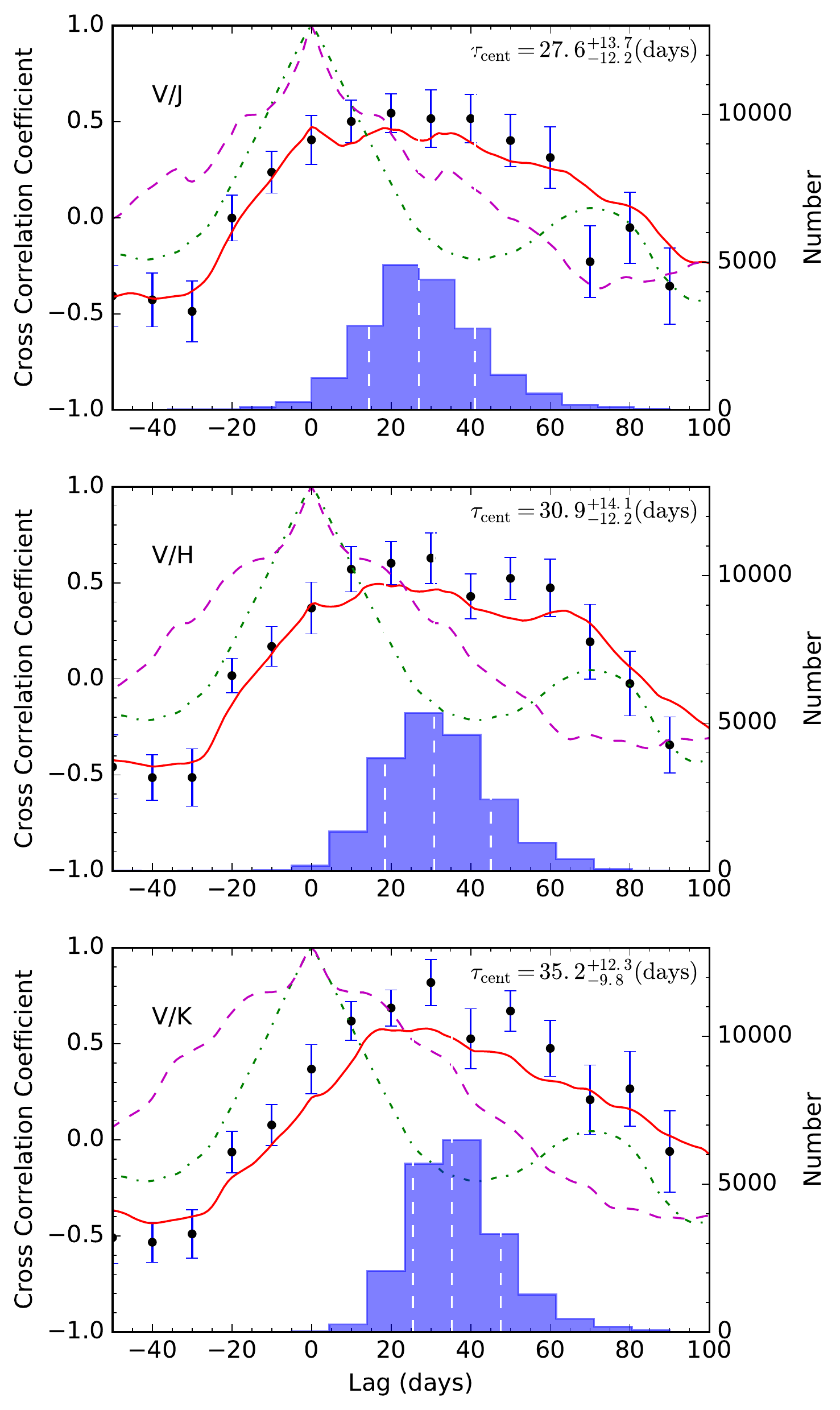}}
\end{center}
\caption{The CCF of $V$ vs. $J$ (top), $V$ vs. $H$ (middle) and $V$ vs. $K$ (bottom) is shown. In each panel, the solid line shows the ICCF, points with error bars show the DCF obtained using  $\Delta \tau = 5$ days and the corresponding distribution of the $\tau_{\mathrm{cent}}$ obtained using 20,000 Monte Carlo Simulations. The dashed-dot line shows the ACF of $V$-band light curve and the dashed line shows the ACF of corresponding NIR band light curve. The value of $\tau_{\mathrm{cent}}$ obtained using Monte Carlo simulation is noted in each panel.}
\label{fig:ccf}
\end{figure}

Our observations indicate that the source H0507+164 show flux variations in the 
optical and NIR bands enabling us to test for the presence/absence of 
time delay between flux variations in different bands. To calculate the time 
lag between the variations in the optical $V$-band emission from the accretion 
disk and the IR emission from the torus, we employed two well-known methods, 
namely the interpolated cross-correlation function 
\citep[ICCF;][]{1986ApJ...305..175G,1987ApJS...65....1G} 
and the discrete correlation function \citep[DCF;][]{1988ApJ...333..646E}. In 
this work, cross-correlation analysis was performed between the $V$-band light curve 
and each of the NIR ($J$, $H$ and $K_s$) light curves.  Figure \ref{fig:ccf} shows 
the result of the cross-correlation analysis between $V$ and $J$ (upper panel),
 $V$ and $H$ (middle panel), $V$ and $K_s$ (lower panel). In each panel, the 
solid line shows the cross-correlation function (CCF) obtained by ICCF method, 
and the data points with error bars are those obtained by DCF method. Both ICCF and 
DCF are found to be in excellent agreement. The autocorrelation function (ACF) 
is also plotted for $V$-band (dashed-dot) and corresponding NIR band (dashed) 
in different panels. The ACFs show zero lag as expected, but the CCFs show an 
overall shift along the positive lag indicating that NIR continuum lags behind the optical $V$-band emission. The amount of lag can be estimated either by the lag corresponding to the peak of the CCF ($\tau_{\mathrm{peak}}$) or by the lag corresponding to the centroid of CCFs ($\tau_{\mathrm{cent}}$). The later has been found to be a better representative of the lag, particularly in 
cases when the light curves are either noisy and/or have less number of points \citep{1998PASP..110..660P}. The centroids of the CCF was calculated as 
\begin{equation}
\tau_{\mathrm{cent}}=\frac{\sum_{i} \tau_{i} \mathrm{CCF}_{i}}{\sum_{i} \mathrm{CCF}_{i}}.
\end{equation}
To quantify the lag, we have used here the centroid of the CCF, which is evaluated by considering all the points that are within 60 per cent of the 
maximum of the CCF. This cut off was selected so as to have sufficient
cross-correlation coefficients for the centroid determination.

The uncertainties in the derived lag were evaluated using a model-independent 
Monte Carlo simulation based on the flux randomization (FR) and random subset 
selection (RSS) described in \citet{1998PASP..110..660P} with the additional 
improvement as suggested by \citet{1999ApJ...526..579W} and summarized by 
\citet{2004ApJ...613..682P}. In each Monte Carlo iteration, we first randomly 
took $N$ independent points from a parent light curve of $N$ data points 
regardless of whether any point has previously been selected. The new light 
curve after RSS method contains $M$ data points.  
%unselected points is $N-M$ in each Monte Carlo iteration. 
To take 
into account the uncertainty in the measured flux values, we then randomly 
modified the fluxes of the $M$ data points by adding the uncertainties of the 
measured flux multiplied with a random Gaussian value. For each Monte Carlo 
iteration, we computed the CCF of the modified light curve and calculated the 
$\tau_{\mathrm{cent}}$ using the points within 60 per cent of the CCF peak. 
This process was repeated for 20,000 iterations retaining only those CCF having
 peak value $>0.5$ so that the cross correlation result is significant. We 
built the cross correlation centroid distribution (CCCD), which is shown by 
the histogram plot in each panel of Figure \ref{fig:ccf}. The median
of the CCCD is taken as a representation of the lag. Since the 
distribution has a non-Gaussian shape, we calculated uncertainties within 68 \% confidence interval around the median value.    %we calculated upper and lower uncertainties in $\tau_{\mathrm{cent}}$ such that 15.87 per cent of the realizations have $\tau>\tau_{\mathrm{cent}}+\delta \tau_{\mathrm{up}}$ and 15.87 per cent have $\tau<\tau_{\mathrm{cent}}-\delta \tau_{\mathrm{low}}$, corresponding to $\pm 1\sigma$ for a Gaussian distribution.

The result of the Monte Carlo simulation using different CCF analysis method 
is given in Table \ref{Table:corr1}. The centroid lag obtained using DCF 
method was calculated for different bin sizes and the results are found to be 
consistent within error bars. The lags obtained by ICCF method are also
given in Table \ref{Table:corr1} which agrees well with the DCF values 
within error bars. Though the time delays 
obtained from both DCF and ICCF agree with each other, for all further 
analysis we consider the lag obtained using the DCF with a bin size of 
$\Delta\tau=5$ days as the typical sampling of
our light curves is about 5 days. Based on this we find the rest
frame (corrected for the redshift) time lags of $27.1^{+13.5}_{-12.0}$ days between 
$V$ and $J$ bands, $30.4^{+13.9}_{-12.0}$ days between $V$ and $H$ bands and $34.6^{+12.1}_{ -9.6}$ days between $V$ and $K_s$ bands. Though there is an indication 
of wavelength dependent lag which increases with wavelength,
because of the larger error bars, we conclude that within errors, the
derived lags are consistent with each other. Considering only $V$ and $K_s$ light curves, 
we obtained a lag of $34.6^{+12.1}_{ -9.6}$ days where $Ks$ band
variation lagging the $V$-band variations. This lag is larger than the lag of 
$3.01^{+0.42}_{-1.84}$ days obtained between the optical and emission line flux variations from 
optical spectroscopic monitoring observations \citep{2011MNRAS.416..225S} and in 
agreement with what is expected from the unification 
model of AGN. %Correcting for time-dilation effects using the redshift of the source we obtained a lag of $34.6^{+12.1}_{-9.6}$ days in the rest frame of the source between V and $Ks$ bands. 
Using this time delay we infer that the inner edge of the 
dusty torus is at a distance of $\approx0.029^{+0.010}_{-0.008}$ pc from the 
central UV/optical AGN continuum source.

\begin{table*}
\centering
\caption{Values of $\tau_{\mathrm{cent}}$ and their associated erros in days as obtained by DCF (for various bin size ($\Delta\tau$)) and ICCF methods.}
\resizebox{0.8\textwidth}{!}{%r
%\hskip-2.0cm
\begin{tabular}{r  r r r r r  r}\hline \hline
%\multicolumn{5}{c}{DCF}
 %Bands &   &   &               CCF Method       &   &   & \\  
  Bands     &   &   & DCF ($\Delta\tau$ in days) &   &   & ICCF \\
       & 2 & 3 & 4   & 5 & 6 & \\\hline
       $V-J$ & $31.1^{+16.5}_{-15.0}$ & $28.6^{+15.4}_{-13.8}$ & $28.0^{+15.1}_{-12.2}$ & $27.6^{+13.7}_{-12.2}$ & $27.3^{+15.1}_{-12.0}$ & $20.7^{+13.5}_{-11.0}$ \\
       $V-H$ & $33.9^{+17.4}_{-14.8}$ & $32.0^{+15.4}_{-13.6}$ & $32.8^{+15.7}_{-14.0}$ & $30.9^{+14.1}_{-12.2}$ & $31.9^{+13.6}_{-12.6}$ & $25.1^{+12.4}_{-11.0}$ \\
       $V-K_s$ & $39.2^{+14.9}_{-12.3}$ & $37.6^{+13.1}_{-11.1}$ & $36.7^{+12.9}_{-10.7}$ & $35.2^{+12.3}_{-9.8}$ & $36.3^{+12.1}_{-10.7}$ & $30.2^{+9.0}_{-7.9}$ \\ \hline

\end{tabular} } %}
\label{Table:corr1}
\end{table*}

\section{Discussion}\label{sec:discussion}
\subsection{Infrared lag and central luminosity correlation}
\begin{figure}
\begin{center}
\resizebox{9cm}{7cm}{\includegraphics{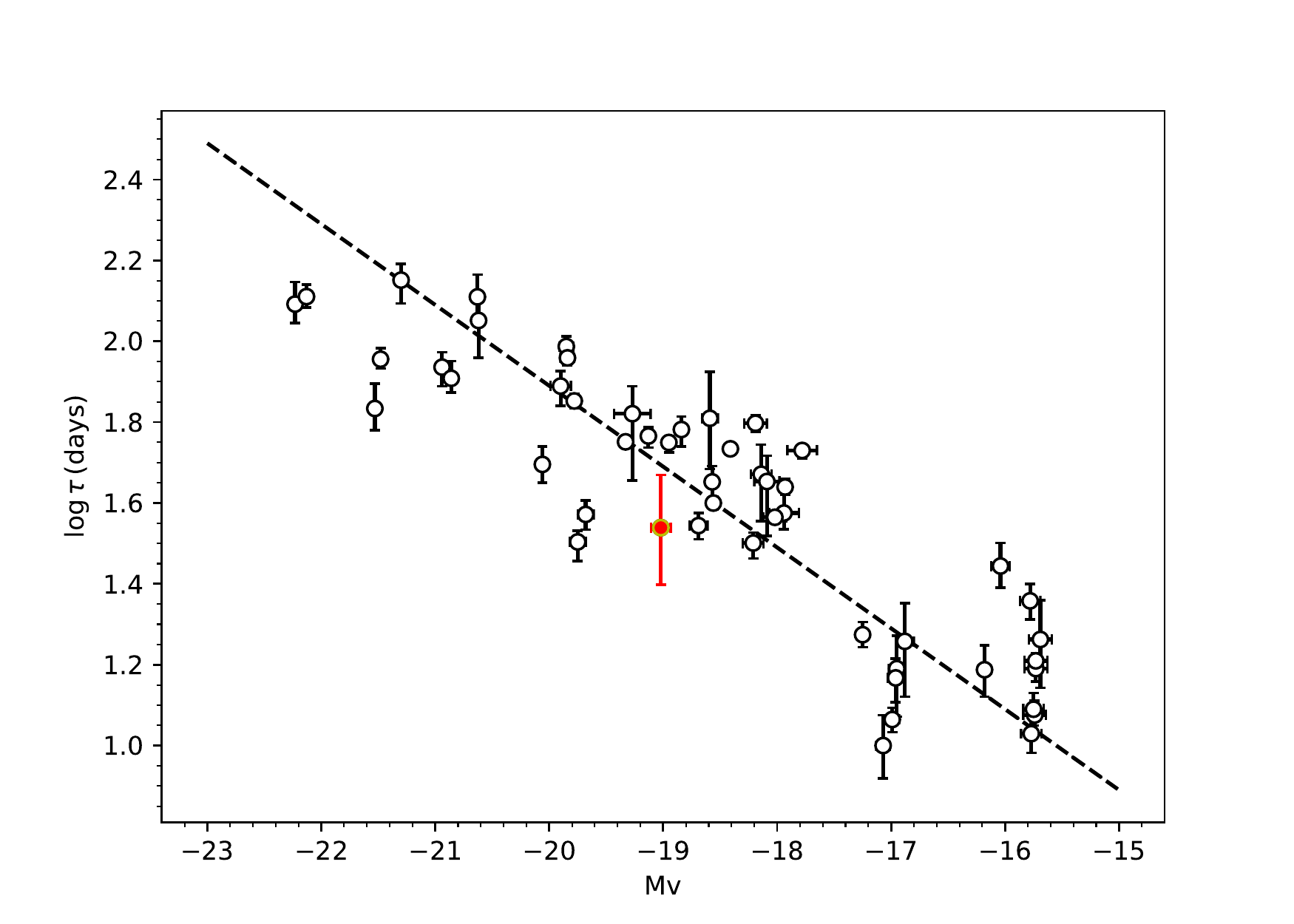}}
\end{center}
\caption{Dust lag-luminosity relationship using the data of \citet{2014ApJ...788..159K}. The filled circle corresponds to the lag between the V and $K_s$ bands determined for} H0507+164 which lies close to the regression line (dashed line), obtained by \citet{2014ApJ...788..159K}. The lag times were corrected for the time dilation effect using the object redshift.
\label{fig:lag-lum}
\end{figure}

Using the rest frame time lag, the inner radius of the dusty torus in H0507+164 was found to be 0.029 pc.
Similar measurements of the radius of the dust torus in about 
two dozen AGN are available in the literature based on DRM observations \citep[see][and the references therein]{2014ApJ...788..159K}. These sources are shown 
in Figure \ref{fig:lag-lum} as empty circles in the lag - luminosity place. 
Also, shown in the same Figure is the 
location of the source H0507+164 indicated by a filled circle.  The best-fitted 
regression line $\log\Delta\tau = -2.11 -0.2 M_V$ of \citet{2014ApJ...788..159K} which
was based on a systematic and homogeneous analysis of DRM data of 17 Seyfert 1 galaxies is shown by a dashed line. Our lag measurement of H0507+164 is 
in excellent agreement with the lag expected for its luminosity and closely 
follows the dust 
lag-optical luminosity correlation of $\Delta\tau_{\mathrm{dust}}\varpropto L^{0.5}$.

\subsection{Structure of the BLR and the Dust Torus}
\citet{2011MNRAS.416..225S} carried out spectroscopic reverberation observations of BLR of H0507+164 and estimated a lag between the optical/UV continuum and the H$\beta$ line emission of $3.01^{+0.42}_{-1.84}$ days in the rest frame of the object. Our DRM observations for the same object in different NIR wavelengths together with the result obtained by \citet{2011MNRAS.416..225S} provide important information about the structure of the BLR and the dust torus of H0507+164. The BLR lag is found to be smaller than the dust-lag (lag between $V$-band and $K_s$-band) which is as expected from the unified scheme of AGN. Though 
there are indications of a wavelength dependent lag in our multi-wavelength 
data reported here, they are not statistically significant due to the large
errors in their lags. From an analysis of a sample of 17 Seyfert galaxies, 
\citet{2014ApJ...788..159K} found that dust reverberation radius of their sample is 4-5 times larger than their BLR radius and typically a 
factor 2 smaller than their interferometric radius. Similarly, the BLR radius observed by reverberation mapping is smaller than the same observed by NIR interferometry \citep[see][]{2012SPIE.8445E..0WP}. Such difference is expected as the reverberation mapping is a response weighted radius originating from the compact region, while the interferometric radius is flux weighted sensitive to the flux coming from the outer region \citep{2007A&A...476..713K}. However, there are exceptions. For the source Mrk 335 \citep{2014ApJ...782...45D} NGC 4151 \citep{2006ApJ...651..775B} and NGC 4593 \citep{2013ApJ...769..128B}, the dust radius \citep{2014ApJ...788..159K} is about 10 times larger than the BLR radius. Such varied differences between the mean dust radius and BLR radius known in AGN will hint for variable dust emission \citep{2009ApJ...700L.109K, 2015A&A...578A..57S}.
To establish this, we  need more precise measurements of dust radius for a large sample of AGN. Our ongoing 
multi-band DRM project on a larger sample will help to understand dust geometry.  

\section{Conclusion}\label{sec:summary}
We have presented the first measurements of the time delays between optical
and NIR bands. This is based on optical $V$ and NIR $J$, $H$ and $K_s$ bands observations spanning a time period of about 170 days. Significant variations have been 
observed in all bands. 
The rest frame lags between $V$-band and NIR bands are found to 
be $27.1^{+13.5}_{-12.0} \, \mathrm{days}$ ($V$ vs. $J$), $30.4^{+13.9}_{-12.0}
 \, \mathrm{days}$ ($V$ vs. $H$) and $34.6^{+12.1}_{-9.6} \, \mathrm{days}$ 
($V$ vs. $K_s$). From the present analysis it is difficult to probe the existence of different lag between $V$  and $J$, $V$ and $H$ and $V$ and $K_s$, solely due to the large errors in the determined lags because of the limited amount of data analysed here. The presence of wavelength dependent lags if any can be established only with the accumulation of good quality monitoring data over a long period of time. 
It is likely that a combined analysis of the data would reduce the error in the lag between optical and NIR \citep{2015A&A...578A..57S}. Given
the limited number of observations in the present work, we have not attempted 
a combined analysis.
 
Using the rest frame lag between $V$ and 
$Ks$-band, we found that the inner radius of the dust torus is 
at a distance of 0.029 pc, from the central UV/optical continuum source.
As expected from the unified model of AGN the dust inner radius is larger than the BLR radius known for this source from spectroscopic reverberation monitoring
observations. Our estimate of  $R_{\mathrm{dust}}$ and the  $V$-band absolute 
magnitude is in good agreement with the dust lag-optical central luminosity 
relationship found by \cite{2014ApJ...788..159K} from a large sample of AGN.

\section*{Acknowledgements}
We thank the referee for his/her valuable comments that helped 
us to improve
our manuscript. S.R. acknowledges the support by the Basic Science
Research Program through the National Research Foundation
of Korea government (2016R1A2B3011457). This research used the SIMBAD data base, which is operated at CDS, Strasbourg, 
France, the NASA/IPAC Extragalactic Database (NED), which is operated by the 
Jet Propulsion Laboratory, California Institute of Technology, under contract
with NASA and data products from the Two Micron All Sky Survey, which is a 
joint project of the University of Massachusetts and the Infrared Processing 
and Analysis Center/California Institute of Technology, funded by the NASA and 
the National Science Foundation. We thank to the supporting staff at the 
Indian Astronomical Observatory (IAO), Hanle, and CREST, Hoskote. AKM and RS 
thank the National Academy of Sciences, India for providing the required fund 
for this project. SH acknowledges support from the European Research Council Horizon 2020 grant DUST-IN-THE-WIND (677117). P.G. thanks STFC for support (grant reference ST/J003697/2). MBP gratefully acknowledges the support from Department of Science and Technology (DST), New Delhi under the INSPIRE faculty  Scheme (sanctioned No: DST/INSPIRE/04/2015/000108).

%\end{twocolumn}

\bibliographystyle{mnras}
\bibliography{ref}

\bsp	% typesetting comment
\label{lastpage}
\end{document}